\documentclass{aa}  

\usepackage{booktabs}
\usepackage{graphicx}
\usepackage{txfonts}
\usepackage{natbib}
\usepackage{placeins}
\bibpunct{(}{)}{;}{a}{}{,} 

\begin{document} 

   \title{LHAASO J1849$-$0002: A hybrid lepto-hadronic interpretation of PeV $\gamma$-ray emission}

\author{Yihan Shi   
        \inst{1} \thanks{These authors contributed equally to this work.}
        \and 
        Yudong Cui
        \inst{1}\footnotemark[1]
        \and
        Lili Yang\inst{1,2}\thanks{Corresponding author: yanglli5@mail.sysu.edu.cn}
}

\institute{School of Physics and Astronomy, Sun Yat-Sen University, No. 2 Daxue Road, Zhuhai 519082, China
\and
Center for Astro-Particle Physics, University of Johannesburg,
P.O. Box 524, Auckland Park 2006, South Africa
}

  \abstract
   {Recently, LHAASO detected $\gamma$-ray emission from the pulsar wind nebula (PWN) J1849-0001 extending up to approximately 2 PeV, providing strong evidence for petaelectronvolt (PeV) particle acceleration. To explain the origin of this ultrahigh-energy emission, we investigated three physical scenarios: a pure leptonic model, a hadronic-dominated model, and a hybrid lepto-hadronic model. We show that while both pure leptonic and hadronic-dominated models can reproduce parts of the multiwavelength spectral energy distribution, neither can simultaneously explain the entire dataset, particularly the PeV tail. The leptonic scenario requires an unrealistically high electron cutoff energy, while the hadronic model underpredicts the highest-energy emission. We therefore propose a hybrid model that combines inverse Compton emission from PWN electrons with hadronic interactions between escaped cosmic rays and a nearby molecular cloud. In this framework, a suppressed diffusion coefficient ($\sim 1\%$ of the Galactic average) is required to confine PeV particles in the source vicinity. This model successfully reproduces the full SED, including the approximately 2 PeV emission. We further calculated the associated neutrino flux and show the sensitivity of NEON to this source. Our results support the interpretation that evolved PWNe embedded in complex environments can act as Galactic PeVatrons.}

   \keywords{Pulsar wind nebula - hybrid model - LHAASO - PeVatron
               }
    \maketitle
\nolinenumbers
\section{Introduction}

The origin of Galactic cosmic rays (CRs) up to the knee at 
approximately petaelectronvolt (PeV) energies remains one of the central open questions in 
high-energy astrophysics. Identifying Galactic ``PeVatrons'', i.e., 
sources capable of accelerating particles to PeV energies, is therefore 
essential for understanding the origin of the highest-energy Galactic CRs. Very-high-energy (VHE; $E \gtrsim 100$~\text{GeV}) and ultrahigh-energy (UHE; $E \gtrsim 100$~\text{TeV}) $\gamma$-ray observations provide a direct way to search for such accelerators because $\gamma$ rays can trace either high-energy electrons through leptonic radiation processes or high-energy hadrons through inelastic interactions with ambient gas. The detection of photons beyond 100 TeV by the Large High Altitude Air Shower Observatory (LHAASO) 
has revealed a population of Galactic UHE $\gamma$-ray sources and demonstrates that PeV particle acceleration can occur in several Galactic environments \citep{Cao2021}. In particular, the High Altitude Water Cherenkov (HAWC) collaboration \citep{abeysekara2020multiple} reports Galactic sources with emission above 56 TeV, including several sources extending beyond 100 TeV, while LHAASO has detected photons up to the PeV range \citep{Cao2021}. Among the possible source classes, pulsar wind nebulae (PWNe) are particularly important candidates, since the rotational energy of young and middle-aged pulsars can power relativistic winds that terminate in shocks and accelerate particles over a broad energy range 
\citep{Gaensler2006}.

The nonthermal emission from PWNe is usually interpreted in terms of relativistic electrons and positrons injected by the pulsar wind. These particles emit synchrotron radiation from radio to X-ray energies in the magnetic field of the nebula and produce gigaelectronvolt to teraelectronvolt (GeV--TeV) $\gamma$ rays through inverse Compton (IC) scattering of ambient photon fields, such as the cosmic microwave background (CMB), infrared radiation, and optical starlight \citep{Gaensler2006, popescu2017radiation}. In evolved systems, the high-energy electrons and positrons may escape from the compact X-ray PWN and diffuse into the surrounding medium, forming an extended TeV halo. Such halos are characterized by $\gamma$-ray emission that is much more extended than the compact X-ray nebula and are generally interpreted as IC emission from escaped or relic electrons around middle-aged pulsars \citep{Abeysekara2017, Linden2017}. Therefore, spatial differences between the X-ray PWN and the TeV $\gamma$-ray emission can provide important information on particle transport, cooling, and the evolutionary stage.

In addition to the leptonic channel, hadronic processes may also contribute to the $\gamma$-ray emission if accelerated CR protons or nuclei interact with dense ambient gas. In this scenario, inelastic proton-proton collisions produce neutral pions that decay into $\gamma$ rays, whereas charged pion and muon decays generate neutrinos \citep{kelner2006energy, Kappes2007}. Molecular clouds located close to a particle accelerator can therefore act as effective targets for escaped CRs, enhancing the hadronic $\gamma$-ray signal and potentially producing a spatial correlation between the $\gamma$-ray morphology and the gas distribution \citep{Gabici2009}. Distinguishing between leptonic and hadronic origins is thus crucial for understanding whether a given UHE $\gamma$-ray source is powered mainly by energetic electrons from a PWN halo, by escaped hadrons interacting with nearby gas, or by a combination of both processes.

PSR J1849-0001 was initially discovered in the X-ray band \citep{terrier2008discovery, gotthelf2011discovery}. It has a spin period of $P$ = 38.5\,ms, a spin-down luminosity of $\dot{E}$ = 9.8 $\times$ 10$^{36}$\,erg\,s$^{-1}$, and a characteristic age of $\tau_c$ = 42.9\,kyr. A fit to the X-ray spectrum yielded a high hydrogen column density of $n_\mathrm{H} = (4.3 \pm 0.6) \times 10^{22}\,\mathrm{cm}^{-2}$, assuming a distance of approximately 7\,kpc \citep{gotthelf2011discovery}.
The best-fit position of LHAASO J1849-0002 is located at R.A. = 282.24\textdegree\ $\pm$ 0.01\textdegree\ and Dec = -0.04\textdegree\ $\pm$ 0.01\textdegree, nearly coincident with the position of PSR J1849-0001, indicating a possible association between LHAASO J1849-0002 and PSR J1849$-$0001 (R.A. = 282.26\textdegree\ and Dec = -0.02\textdegree\ ) \citep{kim2024x,LHAASO2026J1849}. Multiwavelength observations have been obtained near PSR J1849-0001, including X-ray data from XMM-Newton \citep{kim2024x} and Chandra \citep{gagnon2024chandra}, and $\gamma$-ray data  mainly from the High Energy Stereoscopic System (H.E.S.S.) \citep{abdalla2018hess}, the Tibet air shower (AS) and muon detector (MD) array \citep{amenomori2023observation} and LHAASO \citep{Cao2021,LHAASO2026J1849}. \citet{amenomori2023observation} used the Tibet AS and MD arrays to measure the $\gamma$-ray spectrum of H.E.S.S. J1849$-$000 over the 40 to 320 TeV energy range  and report a detection significance of $4.4\sigma$ above 100 TeV. They show that the sub-TeV to sub-PeV spectrum can be reproduced by either a leptonic or a hadronic scenario. In the hadronic interpretation, the inferred proton cutoff energy extends into the PeV range, suggesting that this system may be a Galactic PeVatron. To investigate the origin of the 2 PeV photon, the LHAASO Collaboration recently performed multiwavelength fitting using a leptonic model with a broken power-law electron spectrum, a break at 209.6 TeV, and an average magnetic field of 2.8 $\mu$G \citep{LHAASO2026J1849}. 

In this work, we investigate the origin of the $\gamma$-ray emission from LHAASO J1849$-$0002 by modeling the multiwavelength spectral energy distribution (SED) associated with the PWN of PSR J1849$-$0001. We compare three physical scenarios: a pure leptonic model, a hadronic-dominated model, and a hybrid lepto-hadronic model in which IC emission from PWN electrons is combined with hadronic emission from escaped CRs interacting with a nearby molecular cloud. We further calculate the associated neutrino flux and discuss the prospects for testing the hadronic component with future neutrino observations. The paper is organized as follows. Section 2 summarizes the observational data.  Section \ref{sec:methodology} illustrates the pure leptonic, hadronic-dominated, and hybrid scenarios. We present the results of the three scenarios along with their physical interpretations in Section \ref{sec:discussion}.

\section{Observations} \label{sec:observation}
We used multiwavelength observations from the X-ray to the UHE $\gamma$-ray bands to construct the SED of the PSR J1849$-$0001 region and to investigate the possible spatial association among the PWN, the TeV--PeV $\gamma$-ray emission, and the nearby molecular environment. We analyzed $\gamma$-ray data from H.E.S.S. and LHAASO, which provide complementary coverage in the VHE and UHE regimes.

The H.E.S.S. array consists of imaging atmospheric Cherenkov telescopes located in the Khomas Highland of Namibia. The array detects observed VHE $\gamma$ rays by measuring the Cherenkov light produced by atmospheric particle showers. The current H.E.S.S. system consists of five telescopes and is most sensitive to $\gamma$ rays in the energy range from $\sim 100$ GeV to $\sim 100$ TeV \citep{ohm2023current}. Because of its good angular resolution and sensitivity in the TeV band, H.E.S.S. is particularly useful for resolving extended VHE sources and measuring their spectra. 

The LHAASO observatory is located in Daocheng, Sichuan Province, China, at an altitude of about 4410 m and provides wide-field air-shower observations. It is a multicomponent facility consisting mainly of the Water Cherenkov Detector Array (WCDA), the Kilometer Square Array (KM2A), and the Wide Field-of-view Cherenkov Telescope Array (WFCTA), and is designed for $\gamma$-ray astronomy and CR studies over a broad energy range from the TeV to PeV domain \citep{Cao2019,LHAASO:2019qtb}. Compared with imaging atmospheric Cherenkov telescopes, LHAASO has a wide field of view and a large duty cycle, which makes it well suited for surveying extended $\gamma$-ray sources and detecting UHE photons.

We therefore organize the observational inputs as follows. First, we summarize the X-ray morphology of the PWN as revealed by Chandra and XMM-Newton. We then describe the TeV and PeV $\gamma$-ray observations from H.E.S.S. and LHAASO, respectively. Finally, we examine the CO data around the source to identify possible molecular clouds that may serve as target material for hadronic interactions.

Recently, X-ray observations with Chandra and XMM-Newton have enabled detailed studies of the morphology and spectral properties of the PWN associated with PSR J1849-0001 \citep{kim2024x,gagnon2024chandra}. These studies reveal a compact inner nebula with jet and counter-jet structures, as well as a more extended diffuse nebular component surrounding the pulsar. The extended X-ray emission is spatially associated with the TeV source detected by H.E.S.S. and LHAASO, supporting a possible physical connection between the X-ray PWN and the UHE $\gamma$-ray emission.
Figure 2 of \cite{gagnon2024chandra} shows the X-ray image obtained with XMM-Newton.
\cite{LHAASO2026J1849} show that the region around the pulsar is brighter than the surrounding areas, with a jet feature extending to the northeast and a counter-jet to the southwest. On larger scales, Chandra detected an extended PWN with an elliptical morphology. This region overlaps with the TeV emission region measured by H.E.S.S. and LHAASO. In our work, we define an annular region from $6.6^{\prime\prime}$ to $12.5^{\prime\prime}$ as the inner PWN and a circular region($R < 130^{\prime\prime}$) as the extended PWN. We assume that one group of electrons originates from the inner PWN and another from the halo (extended PWN).

\begin{figure}[htbp]
    \centering
    \includegraphics[width=0.4\textwidth]{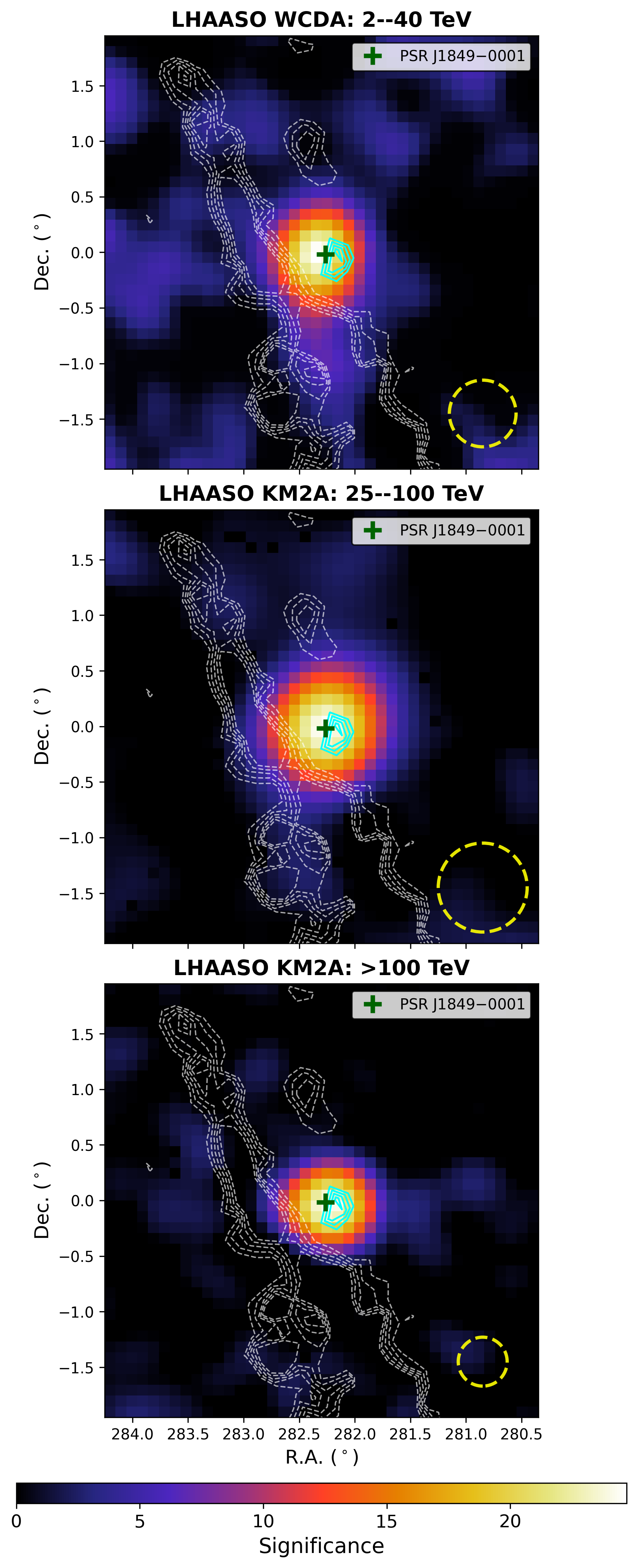}
    \caption{Significance map of LHAASO J1849-0002. 
    Top: WCDA map in the $2 - 20 $ TeV energy range.  
    Middle: KM2A map in the $25 - 100$ TeVe energy range. 
    Bottom: KM2A map at energies $> 100$ TeV. 
    The dark green cross marks the position of PSR J1849-0001. The dashed white lines show significance contours of the molecular clouds from 3.5 K to 12.5 K in steps of 1.5 K, with integration speed ranging from $+94.9 \, \text{km s}^{-1}$ to $+101.4 \, \text{km s}^{-1}$. The cyan lines represent the core cloud. The dashed yellow circles in each panel represent the corresponding angular resolution (PSF) of the observations.}
    \label{fig1}
\end{figure}

At $\gamma$-ray energies, H.E.S.S. detected an extended VHE source
near PSR J1849-0001, with a best-fit angular size of approximately
$0.09^\circ$ \citep{abdalla2018hess}, substantially larger than the
extended X-ray nebula. Subsequent LHAASO observations measured the spectrum and
morphology of J1849-0002 with WCDA and KM2A up to PeV energies (shown in Figure\ref{fig1}),
including photons reaching approximately $2\,\mathrm{PeV}$
\citep{LHAASO2026J1849}. These observations provide
complementary constraints on the source from TeV to PeV energies.

\section{Methodology} 
\label{sec:methodology}
In this work, we model the broadband emission from LHAASO J1849$-$0002 using the radiative code \texttt{GAMERA}\citep{hahn201534th}. We consider three scenarios: a pure leptonic model, a hadronic-dominated model, and a hybrid lepto-hadronic model. This section describes the particle distributions, radiation mechanisms, environmental assumptions, and propagation treatment adopted in each scenario. We present the corresponding fitting results and their physical implications separately in Section\ref{sec:discussion}.

\begin{table}[htbp]
\centering
\caption{Environment parameters of the radiation fields.}
\label{tab:environment}

\begin{tabular}{l c c c}
\toprule
Parameter & Symbol & value \\
\midrule
CMB temperature (K) & $T_{\mathrm{CMB}}$ & 2.73 \\
CMB energy density (erg cm$^{-3}$) & $U_{\mathrm{CMB}}$ & $4 \times 10^{-13}$ \\
far-IR temperature (K) & $T_{\mathrm{fIR}}$ & 30 \\
far-IR energy density (erg cm$^{-3}$) & $U_{\mathrm{fIR}}$ & $2 \times 10^{-12}$ \\
near-IR temperature (K) & $T_{\mathrm{nIR}}$ & 500 \\
near-IR energy density (erg cm$^{-3}$) & $U_{\mathrm{nIR}}$ & $4 \times 10^{-13}$ \\
visible temperature (K) & $T_{\mathrm{SL}}$ & 5000 \\
visible energy density (erg cm$^{-3}$) & $U_{\mathrm{SL}}$ & $1.5 \times 10^{-12}$ \\
\bottomrule
\end{tabular}
\end{table}

\subsection{Leptonic scenario}
\label{subsec:leptonic scenario}
In this scenario, we considered only the contribution from electrons originating from the halo. To calculate the SED of IC scattering by high-energy particles off background soft photons, we adopted the energy densities of the radiation fields obtained from a Galactic radiative transfer model \citep{popescu2017radiation}. 
The radiation fields include the CMB, far-infrared (far-IR), near-infrared (near-IR), and stellar light components, as listed in Table \ref{tab:environment}. We assumed that the electrons follow a simple power-law distribution with an exponential cutoff to describe their maximum energy more precisely, given by 
\begin{equation}
 \frac{dN}{dE} = N_e \left( \frac{E}{1 \, \text{TeV}} \right)^{-\alpha_e}\exp\left(-\frac{E}{E_{cut}}\right).
\label{eq:electron spectrum}
\end{equation}

Here, $\alpha_e$ and $E_{\rm cut}$ are the spectral index and cutoff energy, respectively. The normalization factor, \(N_e\), depends on the total energy \(T_e\) . The relationship between \(N_e\) and \(T_e\) is
\begin{equation}
T_e = N_e \int_{E_{\text{min}}}^{E_{\text{max}}} E \left( \frac{E}{1 \, \text{TeV}} \right)^{-\alpha_e} \exp \left( -\frac{E}{E_{cut}} \right) \, dE.
\label{eq:Te}
\end{equation}

\subsection{Hadronic-dominated scenario} 
\label{subsec: hadronic scenario}
In the leptonic scenario, the same group of electrons that emit IC should also produce synchrotron radiation. Thus, the morphologies of the two emissions are expected to be spatially coincident. However, the observed extents of the X-ray nebula ($2.2'$ \cite{gagnon2024chandra}) and the TeV $\gamma$-ray emission ($5.4'$ \cite{abdalla2018hess}) differ significantly, which cannot be reconciled with a single electron population. This discrepancy motivated us to consider a hadronic scenario. 

The $\gamma$-ray energy spectrum can also be explained within the context of the hadronic scenario, which produces neutrinos. To investigate its feasibility, we examined the CO distribution around the source J1849-0002.
Figure \ref{fig1} shows the outline of the molecular clouds \citep{dame2001milky}. The analysis finds a core cloud of approximately 20 \,pc in size, in the velocity range of 93--100 \, km \, s$^{-1}$, corresponding to a distance of around 7 kpc, which is located on the west side of LHAASO J1849-0002, as shown by the cyan contours in Figure \ref{fig1}. These results are consistent with the previous study by \cite{amenomori2023observation}. The minimum distance between this cloud and the LHAASO source is approximately 17 pc. 
To calculate the mass of the core cloud, we obtained the average molecular hydrogen column density of 
$N(\mathrm{H}_2)\sim1.4\times10^{21}\,\mathrm{cm^{-2}}$, based on its relationship with the integrated line intensity $W(\mathrm{CO})$ and the conversion factor $X_{\mathrm{CO}}$, as in \citep{bolatto2013co}:
\begin{equation}
    N(\mathrm{H_2}) = X_{\mathrm{CO}} \, W(^{12}\mathrm{C}^{16}\mathrm{O} \,~J=1\rightarrow0).
    \label{eq:NH2}
\end{equation}

The conversion factor $X_{\mathrm{CO}} =\allowbreak 2\times10^{20}~\mathrm{cm^{-2}\,(K\,km\,s^{-1})^{-1}}$. The cloud mass, calculated by integrating over the emission region and the velocity range from $+94.9~\mathrm{km\,s^{-1}}$ to $+101.4~\mathrm{km\,s^{-1}}$, is approximately $4.12 \times 10^{4}~M_\odot$. Using the cloud mass, we inferred the approximate number of cold protons, which we used to calculate the production of proton–proton ($pp$) interactions \citep{berezinskii1990cosmic}.

After the escaped CRs propagate to the nearby molecular cloud, they interact with the dense ambient gas and produce a large number of high-energy $\gamma$ rays. The molecular cloud has an average molecular hydrogen column density of $N(\mathrm{H}_2)\sim1.4\times10^{21}\,\mathrm{cm^{-2}}$ . This section discusses the scenario in which the high-energy band is primarily contributed by hadronic processes.

We assumed that the escaped CRs have a simple power-law spectrum of the form
\begin{equation}
 \frac{dN}{dE} = N_p \left( \frac{E}{1 \, \text{TeV}} \right)^{-\alpha_p}\exp\left(-\frac{E}{E_c}\right),
\label{eq:pp SED}
\end{equation}

\begin{equation}
T_p = N_p \int_{E_{\text{min}}}^{E_{\text{max}}} E \left( \frac{E}{1 \, \text{TeV}} \right)^{-\alpha_p} \exp \left( -\frac{E}{E_{c}} \right) \, dE,
\label{eq:Tp}
\end{equation}

where $N_p$ is the normalization factor,  $\alpha_p$ is the spectral index, and $E_{cut}$ is the cutoff energy. Similar to Equation \eqref{eq:Te}, the relationship between the CR proton spectrum and the total energy is given in Equation \eqref{eq:Tp}.
We considered a hadronic model and performed a more detailed
spectral analysis incorporating PWN/SNR evolution, as well as particle propagation in the surrounding environment. The model assumes that escaped CRs from the evolved PWN/SNR  system diffuse into the surrounding interstellar medium and interact with nearby molecular clouds. The same composite environment can also accelerate leptons, producing X-ray nebulae through synchrotron radiation \citep{gelfand2009dynamical,ohira2018pulsar}. In this process, diffusion and particle escape govern the spatial distribution of relativistic protons, while proton--proton interactions with dense gas produce high-energy $\gamma$ rays and neutrinos. In a homogeneous diffusion environment, we employed Green's function \citep{berezinskii1990cosmic,ptuskin2006cosmic} to calculate the propagation of escaped CRs:

\begin{equation}
 G = G(E, r, \Delta t) = \frac{1}{8(\pi \Delta t D)^{3/2}} \exp\left[-\frac{r^2}{4\Delta t D}\right], 
\label{eq:G}
\end{equation}

where $D$ is the diffusion coefficient, $r$ is the distance from the center of the SNR (R.A. = 282.25\textdegree\ and Dec = -0.02\textdegree\ ) to the cloud and \(\Delta t\) is the propagation time. We assumed an average propagation time for all CRs. At \(\Delta t\) ago, most of those CRs were instantly released and started to diffuse. The diffusion coefficient can be written as
\begin{equation}
 D = k D_{10} \left( \frac{E}{10 \, \text{GeV}} \right)^{b}, 
\label{eq:D}
\end{equation}

where \( D_{10} = 1 \times 10^{28} \, \text{cm}^2/\text{s} \), k is the coefficient of $D_{10}$. \cite{berezinskii1990cosmic} and \cite{ptuskin2006cosmic} provide further discussion of these parameters. In addition, we discuss the cloud  masses in Section \ref{sec:observation}.

\subsection{Hybrid scenario} 
\label{subsec:hybrid scenario}

In Sections \ref{subsec:leptonic scenario} and \ref{subsec: hadronic scenario}, we interpret the observational data using pure leptonic and hadronic-dominated models, respectively. This section considers a hybrid model in which leptons are expected to contribute part of the emission in the high-energy band under the condition of a relatively low magnetic field. In the low-energy band, we still adopt the leptonic model described in Section \ref{subsec:leptonic scenario}, with the electron spectrum following Eqs. \ref{eq:electron spectrum} and \ref{eq:Te}.

We also consider the contribution of molecular clouds near the LHAASO source mentioned in Section \ref{sec:observation}. Since the TeV emission is proportional to both the CR density and the target gas density, we expect a good spatial correspondence between the TeV morphology and the ambient gas distribution, as shown in Figure \ref{fig1}.

The $pp$ interaction produces both high-energy $\gamma$ rays and neutrinos through pion and muon decays. Therefore, neutrino detection also serves as a crucial tool for identifying $\gamma$-ray sources. The expected muon neutrino flux can be determined using the following equations \citep{kelner2006energy}:

\begin{equation}
\Phi_{\nu_\mu}(E_{\nu_\mu}) = \frac{c n_{MC}}{4\pi d^2} \int \sigma_{pp}\left(\frac{E_{\nu_\mu}}{x}\right) J_p\left(\frac{E_{\nu_\mu}}{x}\right) F_{\nu_\mu}\left(x, \frac{E_{\nu_\mu}}{x}\right) \, \frac{dx}{x},
\label{eq:neutrino_flux}
\end{equation}

where \(x = E_{\nu_\mu} / E_p\) denotes the integration variable, and \(E_{\nu_\mu}\) and \(E_p\) represent the energy of the neutrino produced and the incident proton, respectively. Here \(c\) is the speed of light, \(n_{MC}\) is the density of the molecular cloud, \(d\) is the distance from PSR J1849-0001 to Earth, for which we adopted $d=7~\mathrm{kpc}$ \citep{gotthelf2011discovery}, and \(J_p\left({E_{\nu_\mu}} / {x}\right)\) denotes the energy distribution of protons as given in Equation \eqref{eq:pp SED}. The inelastic cross section of the $pp$ interaction \(\sigma_{pp}\left({E_{\nu_\mu}} / {x}\right)\) can be written as \citep{kelner2006energy}
\begin{equation}
\sigma_{pp}(E_p) = 34.3 + 1.88L + 0.25L^2 \text{mb},
\label{eq:sigma_pp}
\end{equation}

The function \(F_{\nu_\mu}\left(x, {E_{\nu_\mu}} / {x}\right)\) consists of two components, \(F_{\nu^{(1)}_\mu},F_{\nu^{(2)}_\mu}\) \citep{kelner2006energy}.
 The spectrum of muonic neutrinos produced through the direct decay of pions can be described as follows:

\begin{equation}
\begin{aligned}
F_{\nu^{(1)}_\mu}(x, E_p) 
= {} & B' \frac{\ln(y)}{y}
\left( 
\frac{1 - y^{\beta'}}
{1 + k' y^{\beta'} (1 - y^{\beta'})} 
\right)^4  \\
& \times
\left[
\frac{1}{\ln(y)}
- \frac{4 \beta' y^{\beta'}}{1 - y^{\beta'}}
- \frac{4 k' \beta' y^{\beta'} (1 - 2 y^{\beta'})}
{1 + k' y^{\beta'} (1 - y^{\beta'})}
\right],
\end{aligned}
\label{eq:F_1}
\end{equation}
where $ y = {x}/{0.427} $,
\begin{equation}
B' = 1.75 + 0.204L + 0.010L^2,
\label{eq:B'}
\end{equation}
\begin{equation}
\beta' = \frac{1}{1.67 + 0.111L + 0.0038L^2},
\label{eq:beta'}
\end{equation}
\begin{equation}
k' = 1.07 - 0.086L + 0.002L^2.
\label{eq:k'}
\end{equation}
In the pion decay process, since \(F_{\nu^{(1)}_\mu}\) has a sharp cutoff at $ x = 0.427 $ \citep{kelner2006energy}, the integration range is set from 0 to 0.427. 
The spectrum of muonic neutrinos produced through muon decay is described by

\begin{equation}
F_{\nu^{(2)}_\mu}(x, E_p) = B'' \frac{\left(1 + k'' (\ln x)^2\right)^3}{x \left(1 + 0.3 / x^{\beta''}\right)} (-\ln(x))^5,
\label{eq:F_2}
\end{equation}
where
\begin{equation}
B'' = \frac{1}{69.5 + 2.65L + 0.3L^2},
\label{eq:B''}
\end{equation}
\begin{equation}
\beta'' = \frac{1}{(0.201 + 0.062L + 0.00042L^2)^{1/4}},
\label{eq:beta''}
\end{equation}
and
\begin{equation}
k'' = \frac{0.279 + 0.141L + 0.0172L^2}{0.3 + (2.3 + L)^2}.
\label{eq:k''}
\end{equation}
In the muon decay process, the range of integration is set from 0 to 1.
 
Based on Eqs. \eqref{eq:pp SED}-\eqref{eq:k''}, we calculated the expected muonic neutrino flux from PSR J1849-0001 using the best-fit parameters of our hybrid model, as presented in Section \ref{subsec:hybrid scenario}. The predicted muonic neutrino spectrum is shown by the solid brown line in Figure \ref{fig4}.
We expect that detection of neutrinos can effectively constrain the hadronic contribution of this source. Moreover, we estimated the sensitivity of the Neutrino Observatory Neutrino Observatory in the Nanhai (NEON) \citep{zhang2025proposed} based on our simulation of this source. The predicted neutrino flux is about a factor of two lower than the sensitivity curve.   

\section{Results and discussion} \label{sec:discussion}
In this section, we present the results for the three scenarios described in Section\ref{sec:methodology} and discuss their physical implications. We first investigated whether a single leptonic population could account for the broadband SED. Next, we tested a hadronic scenario in which escaped CRs interact with the nearby molecular cloud. Finally, we assessed a hybrid lepto-hadronic model in which different energy bands arise from different physical regions.

\subsection{Pure leptonic model}
\begin{figure}[htbp]
    \centering
    \includegraphics[width=0.5\textwidth]{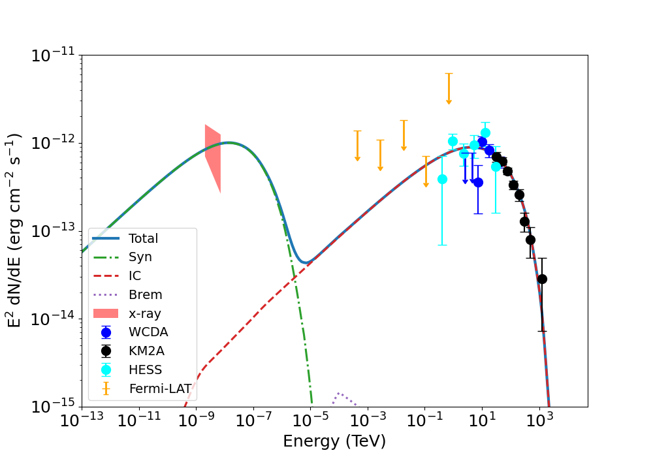}
    \caption{Energy spectrum of PSR J1849-0001 modeled in the leptonic scenario. The shaded red region represents the X-ray flux measured by Chandra \citep{kim2024x}. The yellow upper limits represent GeV $\gamma$-ray flux upper limits at the 95\% confidence level based on Fermi-LAT data \citep{abdollahi2020fermi}. Cyan squares show the H.E.S.S measurements. The blue circles shows WCDA measurements, and black circles shows the KM2A measurements.
    The dash-dotted green, dashed red, and dotted purple lines represent synchrotron, IC, and bremsstrahlung emission, respectively, and the solid blue line represents the total radiation.}
    \label{fig2}
\end{figure}

We explored the main model parameters to identify ranges that are consistent with the observations.
We explored  $E_{cut}$ over the range $500\text{--}800\ \mathrm{TeV}$ because it mainly determines the position and shape of the high-energy spectral cutoff. The $B$ primarily regulates the synchrotron emission and the relative contributions of synchrotron and IC radiation, and we varied it between 2.6 and 3.4 $\mu\mathrm{G}$. The $\alpha_e$ mainly affects the overall spectral slope and shape of the SED, and we considered values in the range 2.3\text{--}2.45. We adjusted the total electron energy accordingly. 
Figure \ref{fig2} shows the best-fit leptonic model, with the parameters \(T_e = 4\times10^{48}\) erg, $\alpha_e = 2.4 $, and $E_{cut} \cong 600$ TeV.
The best-fit magnetic field strength is $B\sim3~\mu\mathrm{G}$. 
This relatively low magnetic field is consistent with previous studies of evolved PWNe, in which the nebular magnetic field is expected to decrease with age \citep{liu2024evolution, manconi2024geminga}.

In the pure leptonic scenario, high-energy electrons in the PWN of PSR J1849-0001 produce the X-ray emission through synchrotron radiation and the TeV–PeV $\gamma$-ray emission through IC scattering of the same electrons on ambient radiation fields. This model can reproduce the broadband SED; therefore, the leptonic interpretation cannot be excluded on spectral grounds alone. Nevertheless, it requires an extremely high electron cutoff energy, $E_{\rm cut}\sim600$ TeV, which is higher than those typically found in PWNe of similar ages \citep{aharonian2006first, aharonian2006energy}. 
In addition, the observed multiwavelength morphology is difficult to reconcile with a single-zone leptonic interpretation. 
The same population of electrons that produces IC $\gamma$ rays should also emit synchrotron radiation in the nebular magnetic field. Therefore, we expect the X-ray and $\gamma$-ray morphologies to be related. However, the observed X-ray nebula has an angular extent of approximately $2.2^{\prime}$ \citep{gagnon2024chandra}, whereas the TeV $\gamma$-ray emission extends to approximately $5.4^{\prime}$ ($0.09^{\circ}$) \citep{abdalla2018hess}. 
This spatial mismatch suggests that the high-energy emission may not arise entirely from the same electron population responsible for the X-ray PWN.

\subsection{Hadronic-dominated model}
\begin{figure}[htbp]
    \centering
    \includegraphics[width=0.5\textwidth]{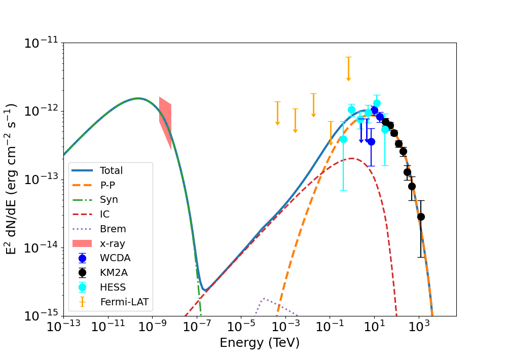}
    \caption{Energy spectrum of PSR J1849-0001 modeled in the hadronic-dominated scenario. The labeled data are the same as in Fig. \ref{fig2}. The long-dashed orange, dash-dotted  green, dashed red, and dotted purple lines represent $pp$ interactions, synchrotron, IC, and bremsstrahlung emission, respectively, and the blue solid line represents the total radiation.}
    \label{fig3}
\end{figure}

Similar to the leptonic model, the range of $\alpha_p$ and $E_c$ are 1.8\text{--}2 and 2\text{--}4 PeV, respectively.  Figure \ref{fig3} shows the best-fit hadronic-dominated model. In this model, we obtain the following CR spectral parameters: 
$T_p = 1 \times 10^{50}~\mathrm{erg}$, $\alpha_p = 1.9$, and $E_c = 3~\mathrm{PeV}$. The best-fit value of the diffusion coefficient are $k = 0.007$ and $b = 0.33$. 
The diffusion distance is 30 pc and the diffusion time is set to $41000~\mathrm{yr}$. For the X-ray observations, we adopted a leptonic model to fit the data. To ensure that the hadronic component dominates the high-energy emission, we suppressed the total leptonic energy ($T_{\rm e}$) by about one order of magnitude relative to the pure leptonic model. The range of $\alpha_e$ and $E_{cut}$ are 1.8\text{--}2.5 and 25\text{--}40 TeV, respectively. The best-fit parameters for the electron spectrum are  \(T_e = 6 \times10^{47}\)erg, $\alpha_e = 2.3 $, and $E_{cut} \cong 30$ TeV. We explored  $B$ values over the range
$13\text{--}19\,\mu\mathrm{G}$, obtaining a best-fit value of
$15\,\mu\mathrm{G}$.

In this scenario, high-energy $\gamma$ rays are produced by $\pi^0$ decay following $pp$ interactions between escaped CR protons and the nearby molecular cloud. The X-ray emission is still described by a leptonic component associated with the PWN. This model naturally explains the extended $\gamma$-ray morphology through the gas distribution, but the magnetic field strength of $15\,\mu$G obtained from our fitting is relatively large, in contrast to the weaker magnetic fields generally expected for older pulsars. We therefore explored a hybrid model that incorporates SNR evolution for a more comprehensive analysis.

\subsection{Hybrid model}
In the hybrid model, the ranges of $\alpha_e$ and $E_{cut}$ are 2.2\text{--}2.4 and 120\text{--}300 TeV, respectively. The best-fit parameters are \(T_e = 1.5\times10^{48}\)erg $\alpha_e = 2.3 $ and $E_{cut} \cong 200$ TeV. We treated the magnetic field strength as a free parameter in the fitting procedure, explored it over the range $3.5\text{--}4.5\,\mu\mathrm{G}$, and obtained a best-fit value of $B = 4~\mu\mathrm{G}$. The ranges of $\alpha_p$ and $E_{c}$ are 1.9\text{--}2.1 and 5\text{--}30 PeV, respectively.
We obtained the following CR spectral parameters: $T_p = 1 \times 10^{50}~\mathrm{erg}$, $\alpha_p = 2$, $E_c = 10~\mathrm{PeV}$. The best-fit diffusion parameters are $k = 0.01$ and $b = 0.33$. The diffusion distance is 40 pc and the diffusion time is set to $41000~\mathrm{yr}$. The diffusion coefficient in this regime is much smaller than the average Galactic value, which is typical of the environments surrounding supernova remnants \citep{semenov2021cosmic}.

In this model, the emission is spatially and physically separated: the lower-energy X-ray emission originates from synchrotron radiation produced by PWN electrons, part of the TeV emission is produced by IC radiation from the leptonic halo, and the highest-energy $\gamma$-ray emission, especially the PeV tail, is mainly attributed to hadronic $pp$ interactions in the nearby molecular cloud illuminated by escaped CRs. We find that the increasing spatial correspondence between the highest-energy LHAASO emission and the molecular cloud supports this hybrid interpretation.

Compared with the pure leptonic scenario, the hybrid model does not require an extremely high electron cutoff energy to reproduce the observed $\sim2$ PeV emission. In addition, unlike the hadronic-dominated model, the hybrid scenario can reproduce the multiwavelength SED with a relatively low magnetic field strength that is more consistent with evolved PWNe. Therefore, the hybrid model provides a more physically self-consistent interpretation of the observational data.

\begin{figure}[htbp]
    \centering
    \includegraphics[width=0.5\textwidth]{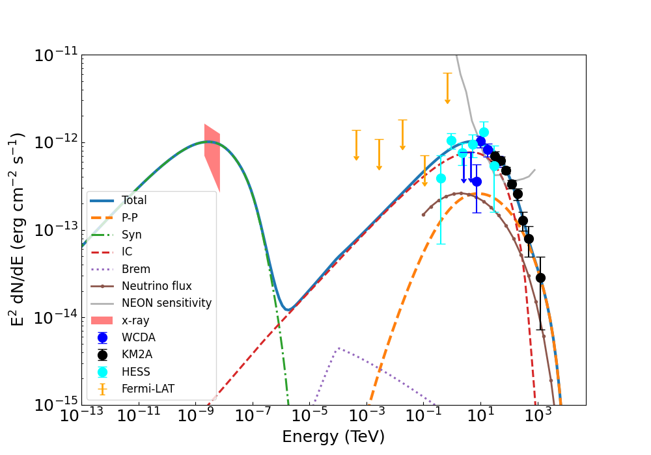}
    \caption{Energy spectrum of PSR J1849-0001 with a hybrid model. The labeled data are the same as in Fig. \ref{fig2}. The long-dashed orange, dash-dotted green, dashed red, and dotted purple lines represent $pp$ interactions, synchrotron, IC, and bremsstrahlung emission, respectively, and the blue solid line represents the total radiation. The solid brown line with dots represents the single-flavor neutrino flux and the gray line represents the NEON sensitivity for this source.}
    \label{fig4}
\end{figure}

Figure \ref{fig4}.  illustrates the role of neutrino observations. The search for Galactic sources constitutes one of the primary scientific objectives of next-generation neutrino observatories. While the estimated sensitivity suggests that a detection remains beyond reach, any future positive signal would nevertheless provide strong constraints on the hadronic component of the hybrid model. In addition, the present calculations already provide valuable input for the experimental design by informing the optimization of the array configuration.

\section{Summary}

This work uses \textsc{GAMERA} \citep{hahn201534th} to explore three scenarios to explain the high-energy observations of LHAASO J1849-0002: a pure leptonic model, a hadronic-dominated model, and a hybrid (lepto-hadronic) model. The first two models can reproduce the observed SED from X rays to $\gamma$ rays, but each has certain limitations.
This spatial correspondence between the higher-energy LHAASO emission and the nearby molecular cloud is naturally expected in the hybrid scenario. We therefore adopt this model because it more naturally accounts for the approximately 2~PeV photons and supports the interpretation that PWNe are viable candidates for PeV particle accelerators.
By combining the spectrum in Figure \ref{fig4} with the image in Figure \ref{fig1}, we find that the spatial correspondence between the LHAASO emission and the nearby molecular cloud becomes more apparent at higher energies. This trend supports a hadronic contribution to the highest-energy $\gamma$-ray emission, particularly the PeV tail, while the X-ray and part of the TeV emission are accounted for by synchrotron and inverse Compton radiation from PWN electrons. The combination of these leptonic and hadronic components therefore favors the hybrid interpretation.

Multiwavelength observations of this region (including neutrino detections) will be crucial to distinguish between leptonic and hadronic dominance and to obtain direct evidence of PeV CR acceleration.

\section*{Data availability}\label{sec:data}

The LHAASO spectral data with error bars presented in Figs. \ref{fig2}, \ref{fig3} and \ref{fig4} are available in electronic form at the Zenodo via \url{https://zenodo.org/records/21672066}.

\begin{acknowledgements}

This work is supported by the National Natural Science Foundation of China (NSFC) grants 12261141691.

\end{acknowledgements}

\bibliographystyle{aa}  
\bibliography{aa}

\end{document}